\documentstyle[12pt]{article}
\begin{document}
\newcommand{\beq}{\begin{equation}}
\newcommand{\eeq}[1]{\label{#1}\end{equation}}
\newcommand{\bea}{\begin{eqnarray}}
\newcommand{\eea}[1]{\label{#1}\end{eqnarray}}
\begin{titlepage}
\begin{center}
\hfill hep-th/9607082\\ \hfill NYU-TH.96/07/01 \\ \hfill CPHT-S460-0796
\vskip .4in
{\large\bf How to Find H-Monopoles in Brane Dynamics}
\end{center}
\vskip .4in
\begin{center}
Massimo Porrati\footnotemark\footnotetext{E-mail: porrati@mafalda.nyu.edu}
\vskip .4in
{\it CPHT, Ecole Polytechnique\\ F-91128 Palaiseau CEDEX, France\\ and\\
Department of Physics, New York University\\
4 Washington Pl., New York NY 10003, USA.\footnotemark}
\footnotetext{Permanent address.}
\end{center}
\vskip .4in
\begin{center} {\bf ABSTRACT} \end{center}
We prove in two different ways that brane dynamics gives rise to all
H-monopoles predicted by S-duality of N=4, 4-dimensional superstring
compactifications. One method uses heterotic-type II duality and the self-dual
3-brane dynamics, the other uses the 5-brane description of small instantons
in $SO(32)$ superstrings.
\begin{quotation}
\noindent
\end{quotation}
\vfill
\end{titlepage}
\noindent
Compactification of the heterotic string on $R^4\times T^6$ yields 12
$U(1)$ fields, $g_{\mu I}$, $B_{\mu I}$, $\mu=0,..,3$, $I=4,..,9$, coming from
the mixed components of the metric and antisymmetric tensor. Elementary
string states with a unit of internal discrete momentum along any of the
compactified dimensions are charged under $g_{\mu I}$, and BPS-saturated
when the Virasoro operator of the
right-moving (supersymmetric) oscillators, $L_0$, vanishes.
These BPS elementary-string states fill 24 short multiplets of the
4-dimensional N=4 supersymmetry\footnotemark.
\footnotetext{More precisely, 21 vector multiplets and one spin-2 short
multiplet.} The model is believed to have
S-duality~\cite{sdual}, this implies that there must exist 24 BPS-saturated
short multiplets with unit magnetic charge with respect to
$B_{\mu I}$~\cite{hmon,h2}: the H-monopoles.
These magnetically charged states are instantons
in the four
coordinates $x^1,x^2,x^3,x^I$, but a field theoretical quantization
of these instantons does not give enough zero modes to produce 24 multiplets.

Purpose of this paper is to show, in two different ways, that all
H-monopoles predicted by S-duality are found, instead, in brane dynamics.
The first one uses the heterotic-type II duality, while the second one proves
their existence directly in the $SO(32)$ heterotic string.

Method 1)

Duality between the N=4, 4-d
heterotic string and the type II string on
$K_3\times T^2$ is well established~\cite{ht2,hs,sen,wit}.
By assuming that the two strings are indeed dual, and using the $O(6,22,Z)$
symmetry of the moduli space of either theory,
one can transform the heterotic gauge fields $g_{\mu I}$, $B_{\mu I}$,
into type IIB gauge fields, $C_\mu$, $D_\mu$,
obtained by dimensional reduction
of the 10-d self-dual 4-form $A$ on $K_3\times T^2$~\cite{ver,gp}:
\beq
C = \int A \wedge (\gamma \wedge a), \;\;\; D=\int A \wedge (\gamma^*\wedge a).
\eeq{-1}
Here, $\gamma$ is a supersymmetric
2-cycle of $K_3$, $\gamma^*$ is its dual, while $a$ is a 1-cycle of $T^2$.

Electrically charged states of $C_\mu$, and magnetically charged states of
$D_\mu$, appear symmetrically:
the former are given by the 10-d, self-dual
3-brane wrapped on $\gamma\wedge a$, while the
latter are given by the self-dual 3-brane wrapped on $\gamma^*\wedge a^*$.
The Narain lattice of the (6-d) heterotic string, $\Gamma_{(4,20)}$,
is mapped by duality into the cohomology lattice $H_*(K_3,Z)$~\cite{wit},
and the lorentzian norm of the former is mapped into the intersection form
of the latter.
Because of this identification, states with a unit of electric
(magnetic) charge, are given by the 3-brane wrapped around $\gamma_0\wedge a$
($\gamma_0^*\wedge a^*$), where $\gamma_0$ is a supersymmetric
2-cycle of $K_3$ with zero self-intersection number.
The results of ref.~\cite{vafa} ensure that this
2-cycle can be realized by a genus-1 holomorphic curve in $K_3$, and that
the dynamics of the 3-brane on $\gamma_0\wedge a$ ($\gamma_0^*\wedge a^*$)
is given by a supersymmetric sigma model on $K_3$~\cite{vafa}.
Each vacuum of this sigma model gives rise to a BPS saturated multiplet.
The degeneracy of the ground state is $\mbox{dim}\,H_*(K_3)=24$, which gives
exactly 24 electrically (magnetically) charged short multiplets of N=4!
Notice that the 3-brane description of H-monopoles is manifestly
S-dual, thus, proving the existence of H-monopoles is no different than
proving the existence of electrically charged states.

Method 2)

To solve the H-monopole problem directly in the heterotic string is less easy.
A major step towards the resolution of the puzzle was made in
ref.~\cite{si}. There, it was discovered that small instantons in the $SO(32)$
heterotic string are described by a 5-brane with N=1 world-volume
supersymmetry. Small instantons are relevant in the H-monopole problem because
in a generic N=4 vacuum the unbroken gauge group of the string
compactification is abelian, and instantons classically shrink to a point.

Reference~\cite{si} also contains a conjecture on how to find the missing
H-monopoles by quantization of the 5-brane.
In the rest of the paper, we show that that conjecture works, with only
relatively minor modifications.

The $SO(32)$ 5-brane contains several 6-dimensional N=1
multiplets~\cite{si}:
an $SU(2)$ vector multiplet, a neutral hypermultiplet, and  charged
hypermultiplets transforming as $(32,2)$ of $SO(32)\times SU(2)$.
The BPS saturated states are the ground states of the quantized 5-brane.
Quantization of the (free) hypermultiplet gives rise to a single N=4 short
multiplet, which has unit magnetic charge under $B_{\mu I}$ if the 5-brane
wraps once around all compactified dimensions except $I$. From now on we shall
set $I=4$. The multiplicity of
BPS multiplets is given, therefore, by the number of ground states of the
6-dimensional N=1 supersymmetric theory compactified on a 5-torus (i.e. in
a periodic box) and made of the remaining fields. Since the Wilson lines of
$SO(32)$ do not fluctuate and appear as coupling constants in the world-volume
theory, this theory describes  32 massive hypermultiplets,
doublets of $SU(2)$, interacting with an $SU(2)$ (triplet) vector multiplet.

Let us review the analysis of the quantized 5-brane ground states
made in ref.~\cite{si}.

Because of the Wilson lines of $SO(32)$, the hypermultiplets have no zero
modes, and in first approximation they can be ignored. The zero modes of the
$SU(2)$ vector multiplet are spanned by commuting Wilson line that can be
taken to be $W_r=\mbox{diag}(e^{2i\pi a_r}, e^{-2i\pi a_r})$, $r=5,..,9$,
which means that the moduli space of the vector multiplet is $T^5/Z_2$.
The $a_r$s
and their supersymmetric partners (four complex Fermi variables) are odd
under the $Z_2$ Weyl symmetry, and their dynamics is free. The quantization of
the bosons $a_r$ gives a unique ground state wave function $\psi(a_r)=1$,
while the quantization of the Fermi variables gives 16 degenerate states,
eight
of which survive the Weyl projection: the naive quantization of the 5-brane
gives eight of the 24 BPS states we are looking for.
The remaining 16 multiplets should come from the 16 singular points in
$T^5/Z_2$ where
the Wilson lines of $SU(2)$ cancel the effect of the Wilson lines of $SO(32)$
in the charged hypermultiplets, so that they may become relevant in the
low-energy dynamics of the 5-brane. Ref.~\cite{si} suggests that it should be
possible to
study the effect of these singularities locally around each one
of them. In the neighborhood of any of the 16 singularities,
the low-energy dynamics of
the 5-brane is given by dimensional reduction to zero space dimensions of
a 6-dimensional N=1 $U(1)$ gauge theory with a charged hypermultiplet.
A normalizable bound state of the resulting supersymmetric quantum mechanics
would give rise to an extra vector multiplet for each singularity.

Unfortunately, the extra ground state associated with the singularity
is not easily localized around it. The reason is that the supersymmetric
quantum-mechanical
hamiltonian  describing the dimensionally reduced $U(1)$ gauge
theory and the hypermultiplet has a homogeneous potential, with a gapless
spectrum. This means that the ground-state wave function --if it exists at
all-- can decay at most
polynomially in the distance from the singularity, making a direct 
study of the problem quite tricky.
In spite of this,
we shall see that the extra short multiplets we look for are indeed associated
with the singularities, as conjectured in ref.~\cite{si}, even though in a less
direct way.

Our strategy for studying the quantization of the 5-brane is the following:
first, we compactify the 5-brane world-volume lagrangian to two (space-time)
dimensions on $T^4$; then, we add to the
 2-dimensional lagrangian a deformation which does not change the number of
its ground states, and which makes it possible to compute their degeneracy
semiclassically.

Notice that the moduli space of the 5-brane is compact ($T^5/Z_2$). In order to
preserve this compactness, we cannot just perform a dimensional reduction to
two dimensions, rather, we must keep also those Kaluza-Klein states that may
become massless at the 16 special points in $T^5/Z_2$~\footnotemark.
\footnotetext{The singularity at $W_r=\pm 1$ is harmless, and its only effect
is to project out states odd under the Weyl projection~\cite{wi}.}
The world-volume action of the 5-brane is uniquely fixed by supersymmetry up
to two derivative terms; we refer to~\cite{t} for the explicit form of a
generic 6-dimensional supersymmetric action.
Upon compactification on $T^4$ to
two dimensions, and after eliminating all Kaluza-Klein modes that {\em never}
develop zero modes on the moduli space --which are, therefore, irrelevant in
the study of the ground states-- we find a 2-dimensional sigma
model with N=4, 2-d supersymmetry. It can be written in terms of 2-d
superfields in a way that makes manifest one of the four 2-d supersymmetries.
We shall use the notations and conventions of ref.~\cite{f}.
Dimensional reduction of the hypermultiplets gives rise to the following
2-d scalar superfields~\cite{t,f}:
\beq
\Phi^{aA}_{i {\bf n}}, \;\;\;
\Phi^{aA}_{i {\bf n}} =\varepsilon_{ij}\varepsilon^{ab}
(\Phi^{bA}_{j {\bf n}})^*.
\eeq{3}
Here $a=1,2$ is the gauge $SU(2)$ index, $A=1,..,32$ is the $SO(32)$ index,
$i=1,2$ is the global $SU(2)$ index necessary to define Majorana spinors in
six dimensions~\cite{t}, and ${\bf n}\equiv(n,\vec{n})\in Z^4$ labels the
Kaluza-Klein tower of states of the 6-d hypermultiplet.
Three additional scalar superfields come from the dimensional reduction of
the 6-d $SU(2)$ vector multiplet. Since we are interested only in the moduli
dynamics, we keep only those superfields whose first component are the moduli
$a_r$,
$r=7,8,9$, and we denote them by~\cite{f}
\beq
\vec{A}=\vec{a}+ i\theta\vec{\lambda} +{i\over 2}\bar{\theta}\theta \vec{F}.
\eeq{4}
The modulus $a_6$, instead, together with the 2-d vector,
belongs to a 2-d vector superfield which reads~\cite{f}:
\beq
V_\alpha=\xi_\alpha +{1\over 2} A^\mu(\gamma_5\gamma_\mu\theta)_\alpha +
{1\over 2}a_6 \theta_\alpha +{1\over 2}N(\gamma_5\theta)_\alpha +{i\over 2}
\bar{\theta}\theta\zeta_\alpha.
\eeq{5}
To write a manifestly N=1 action is now almost immediate, using the results of
ref.~\cite{f}. The only subtlety is that we must make sure that the action
respects the periodicity $a_r \sim a_r +2\pi n_r$, inherited from 6-d, as well
as the Weyl symmetry.
This can be done by introducing a tower of auxiliary vector superfields
$V_\alpha^n$ and a scalar-superfield Lagrange multiplier, $\Lambda^n$, which
implements the constraint
\beq
V_\alpha^n = V_\alpha + {1\over 2}n\theta_\alpha +
i(\gamma_5 D)_\alpha \Omega^n,
\eeq{7}
where the $\Omega^n$s are scalar superfields.
Upon introduction of the covariant spinor derivative $\nabla_\alpha^n \equiv
D_\alpha + (\gamma_5 V^n)_\alpha(\sigma_3)_a^b$, the 2-d action reads~\cite{f}:
\bea
S&=&\int d^2\theta d^2x \Big[ {1\over 2}\bar{D}V \bar{D}D
\bar{D}V -{1\over 2} \bar{D} \vec{A}\cdot D \vec{A}
- {1\over 2}
\sum_{\bf n}(\nabla^{n} \Phi)^{*aA}_{i{\bf n} }(\nabla^n\Phi)^{aA}_
{i{\bf n}} + \nonumber \\
& & + \sum_n \Lambda^n (DV^n - DV^n - n) +
V(\vec{A}, \Phi^{aA}_{i{\bf n}})\Big].
\eea{8}
The {\em real} superpotential $V(\vec{A}, \Phi^{aA}_{i{\bf n}})$ is easily
determined by dimensional reduction of the 6-d N=1 supersymmetric action with
a modulus vector multiplet and the hypermultiplets~\cite{t,f}
\beq
V(\vec{A}, \Phi^{aA}_{i{\bf n}})={1\over 2}\sum_{\bf n}\Big[
\Phi^{aA}_{i{\bf n}}(\vec{A}-\vec{n})\cdot(\varepsilon \vec{\sigma})_{ij}
(\varepsilon\sigma_3)_{ab}\Phi^{bA}_{j{\bf n}}  +
\Phi^{aA}_{i{\bf n}}\vec{\Sigma}_{AB} \cdot(\varepsilon\vec{\sigma})_{ij}
\varepsilon_{ab}\Phi^{bB}_{j{\bf n}}\Big].
\eeq{9}
Here, $\vec{\Sigma}_{AB}=-\vec{\Sigma}_{BA}$ are the Wilson lines of $SO(32)$
along $T^3$.
The Weyl projection further implies that the ground states of the
5-brane are even under the parity
\beq
\vec{A}\rightarrow -\vec{A},\;\;\; V_\alpha \rightarrow -V_\alpha ,\;\;\;
\Phi^{aA}_{i{\bf n}}\rightarrow \varepsilon_{ab}\Phi^{bA}_{i{\bf -n}}.
\eeq{10}

The superpotential in eq.~(\ref{9}) is N=4 supersymmetric in 2-d. By adding
to it {\em any} smooth, even, real function of $\vec{A}$,
$\delta V(\vec{A})$, periodic under
$\vec{A}\rightarrow \vec{A} + \vec{n}\in Z$, we break N=4 but preserve N=1.
Compactness of $T^3$ (the moduli space spanned by $\vec{A}$) implies that
the Witten index~\cite{wi} of the theory is not modified by the perturbation.

To know the number of ground states
--and not just the difference between bosonic and fermionic ones, given by
the index-- we need a perturbation $\delta V$ which allows us to compute the
index semiclassically, and which gives rise to no fermionic vacua.
In this case, indeed, one can argue as follows: to each (normalizable)
ground state of the unperturbed theory, $\Psi$, one can associate a state
$\exp[\delta V(\vec{A})]\Psi$, which is also normalizable since the moduli
space $T^3$ is compact. This state is cohomologically equivalent
to a ground state of the perturbed theory~\cite{wi}.
Since no fermionic vacua exist in
the perturbed theory, the unperturbed theory has no
fermionic vacua either, and therefore the index counts the total number of
(bosonic) vacua.

A convenient choice for $\delta V$, satisfying these criteria, as we shall see
in a moment, is
\beq
\delta V(\vec{A})= {\mu \over 2\pi}\sum_{r=7}^9 \cos 2\pi A_r,\;\;\;
\mbox{$\mu=$ nonzero constant}.
\eeq{11}
The index is given by finding the semiclassical vacua around the
stationary points of $V+\delta V$:
\bea
{\partial\over \partial \Phi^{aA}_{i{\bf n}}}(V+\delta V)&=&
(\vec{A}-\vec{n})\cdot(\varepsilon \vec{\sigma})_{ij}
(\varepsilon\sigma_3)_{ab}\Phi^{bA}_{j{\bf n}} +
\vec{\Sigma}_{AB}\cdot (\varepsilon\vec{\sigma})_{ij}
\varepsilon_{ab}\Phi^{bB}_{j{\bf n}}=0,     \nonumber \\
{\partial\over \partial A_r}(V+\delta V)&=&
{1\over 2}\sum_{\bf n}
\Phi^{aA}_{i{\bf n}}(\varepsilon \sigma^{r-6})_{ij}
(\varepsilon\sigma_3)_{ab}\Phi^{bA}_{j{\bf n}} -\mu \sin 2\pi A_r=0.
\eea{12}
They fall into two classes\footnotemark.
\footnotetext{By symmetry, the range of $A_r$ can be restricted to
$0\leq A_r <1$.}

A) $A_r=0, 1/2,$
$\Phi^{aA}_{i{\bf n}}=0 \; \forall {\bf n}.$

Semiclassical expansion around these $2^3=8$ points gives rise to eight bosonic
plus eight fermionic vacua, because at these points there is single
2-d Majorana-fermion zero mode $\lambda_\alpha=\zeta_\alpha -
(\gamma \partial \xi)_\alpha$, superpartner of $a_6$ and $A^\mu$.
The Weyl projection~(\ref{10}) eliminates the eight fermionic vacua.
The surviving ones are in one-to-one correspondence with the vacua found by
quantizing the vector multiplet alone~\cite{si}.

B) $\det[(\vec{A}-\vec{n})\cdot(\varepsilon \vec{\sigma})_{ij}
(\varepsilon\sigma_3)_{ab}\delta_{AB} + \vec{\Sigma}_{AB}\cdot
(\varepsilon\vec{\sigma})_{ij}\varepsilon_{ab}]=0,$ for some $\vec{n}$,

$\;\;\;\;\;(1/2)\sum_{\bf n}
\Phi^{aA}_{i{\bf n}}(\varepsilon \sigma^{r-6})_{ij}
(\varepsilon\sigma_3)_{ab}\Phi^{bA}_{j{\bf n}} -\mu \sin 2\pi A_r=0.$

These 32 points are symmetric under $\vec{A}\rightarrow -\vec{A}$.
For a generic value of the $SO(32)$ Wilson lines
$\vec{\Sigma}_{AB}$, a single hypermultiplet gets a nonzero expectation value.
We are in the Higgs phase: the
gauge group is completely broken, and all 2-d fermions become massive and
have no zero modes. A short calculation shows that the fermion mass matrix has
positive determinant around all stationary points;
therefore, each of them gives a single bosonic vacuum.
The Weyl projection eliminates half of these states. The surviving ones are
in one-to-one correspondence with 16 vacua of the unperturbed theory, which
give rise to 16 BPS saturated N=4 vector multiplet. These are the missing
H-monopoles. As conjectured in ref.~\cite{si}, they are associated with the
16 singularities in moduli space where some hypermultiplet becomes massless,
even though the correspondence found here is less direct that in~\cite{si}.

Differently from our previous method,
we do not have a physical picture of the additional ground states in terms of
D-brane dynamics.

Either of the two methods developed here might have wider
applications, for instance in finding the multiplicity of states that
preserve only $1/4$ of the supersymmetry.
This problem is related to the attempt of finding a microscopic derivation of
the Bekenstein-Hawking entropy of a black hole~\cite{sv}, and has recently
been addressed in~\cite{v}.
\vskip .1in
\noindent
Acknowledgements
\vskip .1in
\noindent
This work was supported in part by NSF Grant PHY-9318781.


\begin{thebibliography}{99999}
\bibitem{sdual} J.H. Schwarz and A. Sen , Phys. Lett. B312 (1993) 105,
hep-th/9305185.
\bibitem{hmon} J. Gauntlett and J.A. Harvey, hep-th/9407111.
\bibitem{h2} L. Girardello, M. Porrati and A. Zaffaroni, hep-th/9508056,
to appear in Int. Jou. of Mod. Phys. A.
\bibitem{ht2} C.M. Hull and P.K. Townswend, Nucl. Phys. B438 (1995) 109,
hep-th/9410167.
\bibitem{hs} J.A. Harvey and A. Strominger, Nucl. Phys. B449 (1995) 535,
hep-th/9504047.
\bibitem{sen} A. Sen, Nucl. Phys. B450 (1995) 103, hep-th/9504027.
\bibitem{wit} E. Witten, Nucl. Phys. B443 (1995) 85 , hep-th/9503124;
hep-th/9507121.
\bibitem{ver} E. Verlinde, Nucl. Phys. B455 (1995) 211, hep-th/9506011.
\bibitem{gp} A. Giveon and M. Porrati, hep-th/9605118, to appear in Phys.
Lett. B.
\bibitem{vafa} M. Bershadsky, V. Sadov and C. Vafa, Nucl. Phys. B463 (1996)
420, hep-th/9511222.
\bibitem{si} E. Witten, Nucl. Phys. B460 (1996) 541, hep-th/9511030.
\bibitem{wi} E. Witten, Nucl. Phys. B202 (1982) 253.
\bibitem{t} P.S. Howe, G. Sierra and P.K. Townsend, Nucl. Phys. B221 (1983)
331; G. Sierra and P.K. Townsend, Nucl. Phys. B233 (1983) 289.
\bibitem{f} S. Ferrara, Nuov. Cim. Lett. 13 (1975) 629.
\bibitem{sv} A. Strominger and C. Vafa, hep-th/9601029.
\bibitem{v} R. Dijkgraaf, E. Verlinde and H. Verlinde, hep-th/9607026.
\end{thebibliography}
\end{document}